\documentclass{article}
\usepackage{amssymb}
\usepackage{bm}
\usepackage{booktabs}
\usepackage{graphicx}
\usepackage[hmargin=3cm,vmargin=3.5cm]{geometry}
\def\T{{ \mathrm{\scriptscriptstyle T} }}
\begin{document}
\title{A simple sampler for the horseshoe estimator}%
\author{Enes Makalic \and Daniel F. Schmidt}%
\maketitle
\abstract{In this note we derive a simple Bayesian sampler for linear regression with the horseshoe hierarchy. A new interpretation of the horseshoe model is presented, and extensions to logistic regression, negative-binomial regression and alternative prior hierarchies, such as horseshoe$+$, are discussed. Due to the conjugacy of the proposed hierarchy, Chib's algorithm may be used to easily compute the marginal likelihood of the model.}
%
\section{Introduction}
Consider the following Bayesian linear regression model for data $y \in \mathbb{R}^n$
\begin{eqnarray}
		{\bf y} | {\bf X}, \bm{\beta}, \sigma^2 &\sim& \mathcal{N}_n ({\bf X} \bm{\beta}, \sigma^2 {\bf I}_n), \label{eqn:y} \\
		\beta_j | \lambda_j^2, \tau^2, \sigma^2 &\sim& \mathcal{N}(0, \lambda_j^2 \tau^2 \sigma^2), \\
		\sigma^2 &\sim& \sigma^{-2} d\sigma^2, \\
		\lambda_j  &\sim& \mathcal{C}^+(0, 1), \\
		\tau &\sim& \mathcal{C}^+(0, 1) \label{eqn:sigma2} 
\end{eqnarray}
where ${\bf X} \in \mathbb{R}^{n \times p}$ is a matrix of predictor variables (not necessarily full rank), $\mathcal{N}_k(\cdot,\cdot)$ is the $k$-variate Gaussian distribution, $\mathcal{C}^+(0, 1)$ is the standard half-Cauchy distribution with probability density function
\begin{equation}
	p(z) = \frac{2}{\pi(1 + z^2)}, \quad z > 0, \nonumber
\end{equation}
and $j = (1,\ldots,p)$. It is usual to require that the $p$ predictors are standardized to have zero mean and unit length and that the data $y$ is centred. This avoids the need to explicitly model a separate parameter for the intercept.

Equations (\ref{eqn:y}--\ref{eqn:sigma2}) define the horseshoe regression hierarchy recently proposed in~\cite{CarvalhoPolson10}. The horseshoe model is a global-local shrinkage procedure in which the local shrinkage for coefficient $\beta_j$ is determined by $\lambda_j > 0$ and the overall level of shrinkage is determined by the hyperparameter $\tau > 0$. The particular choice of a half-Cauchy prior distribution over the global and local hyperparameters results in aggressive shrinkage of small coefficients (i.e., noise) and virtually no shrinkage of sufficiently large coefficients (i.e., signal). This is in contrast to the well known Bayesian lasso~\cite{ParkCasella08} and Bayesian ridge hierarchies where the shrinkage effect is uniform across all coefficients. Further favourable properties of the horseshoe model are discussed in \cite{CarvalhoPolson10,PolsonScott12a,PolsonScott10}.

The original horseshoe paper does not provide details for efficient sampling from the posterior distribution of the regression coefficients. A standard Gibbs sampling approach is difficult to implement due to the non-standard form of the conditional posterior distributions for the hyperparameters $(\lambda_1,\ldots,\lambda_p)$ and $\tau$. Subsequent papers have suggested the use specialised algorithms, such as slice sampling~\cite{Neal03}, for the hyperparameters~\cite{PolsonEtAl2014}. In this paper, we provide an alternative sampling scheme for all model parameters based on auxiliary variables that leads to conjugate conditional posterior distributions for all parameters, making the application of Gibbs sampling relatively straightforward.  
\section{Bayesian horseshoe with auxiliary variables}
We make use of the following scale mixture representation of the half-Cauchy distribution. Let $x$ and $a$ be random variables such that
\begin{equation}
\label{eqn:half-cauchy}
	x^2 | a \sim \mathcal{IG}(1/2, 1/a) \quad {\rm and} \quad a \sim \mathcal{IG}(1/2, 1/A^2);
\end{equation}
then $x \sim \mathcal{C}^+ (0, A)$~\cite{WandEtAl11}, where $\mathcal{IG}(\cdot,\cdot)$ is the inverse-gamma distribution with probability density function
\begin{equation}
	p(z | \alpha, \beta) = \frac{\beta^\alpha}{\Gamma(\alpha)} z^{-\alpha-1} \exp \left( -\frac{\beta}{z} \right). \nonumber
\end{equation}
Using the decomposition (\ref{eqn:half-cauchy}) leads to the revised horseshoe hierarchy
\begin{eqnarray*}
		{\bf y} | {\bf X}, \bm{\beta}, \sigma^2 &\sim& \mathcal{N}_n ({\bf X} \bm{\beta}, \sigma^2 {\bf I}_n), \\
		\beta_j | \lambda_j^2, \tau^2, \sigma^2 &\sim& \mathcal{N}(0, \lambda_j^2 \tau^2 \sigma^2), \\
		\sigma^2 &\sim& \sigma^{-2} d\sigma^2, \\
		\lambda^2_j | \nu_j &\sim& \mathcal{IG}(1/2, 1/\nu_j), \\
		\tau^2 | \xi &\sim& \mathcal{IG}(1/2, 1/\xi), \\
		\nu_1,\ldots,\nu_p, \xi &\sim& \mathcal{IG}(1/2, 1). 
\end{eqnarray*}
The above hierarchy makes Gibbs sampling from the posterior distribution straightforward. The conditional posterior distribution of the regression coefficients $\bm{\beta} \in \mathbb{R}^p$~\cite{LindleySmith72} is
\begin{equation}
\label{eqn:beta:condp}
	\bm{\beta} | \cdot \sim \mathcal{N}_p ( {\bf A}^{-1} {\bf X}^{\T} {\bf y}, \sigma^2 {\bf A}^{-1} ), \quad {\bf A} = ({\bf X}^{\T} {\bf X} + \bm{\Lambda}_*^{-1}), \quad \bm{\Lambda}_* = \tau^2 \bm{\Lambda},
\end{equation}
where $\bm{\Lambda} = {\rm diag}(\lambda_1^2,\ldots,\lambda_p^2)$. The conditional posterior distribution of $\sigma^2$ is an inverse-gamma distribution given by
\begin{equation}
	\sigma^2 | \cdot \sim \mathcal{IG}\left( (n + p)/2, ({\bf y} - {\bf X}\bm{\beta})^{\T}({\bf y} - {\bf X}\bm{\beta})/2 + \bm{\beta}^{\T} \bm{\Lambda}^{-1}_* \bm{\beta} / 2 \right). \nonumber
\end{equation}
The conditional posterior distributions for the local and global hypervariances are also of inverse-gamma type
\begin{eqnarray}
	\lambda_j^2 | \cdot &\sim& \mathcal{IG} \left(1, \frac{1}{\nu_j} + \frac{ \beta_j^2}{2 \tau^2 \sigma^2}  \right), \quad (j=1,2,\ldots,p), \label{eqn:hs:lam2} \\
	\tau^2 | \cdot &\sim& \mathcal{IG} \left(\frac{p+1}{2}, \frac{1}{\xi}  + \frac{1}{2\sigma^2} \sum_{j=1}^p \frac{\beta_j^2}{\lambda_j^2 } \right). \nonumber
\end{eqnarray}
Finally, the conditional posterior distributions for the auxiliary variables are:
\begin{eqnarray*}
	\nu_j | \cdot &\sim& \mathcal{IG} \left(1, 1 + \frac{1}{\lambda_j^2} \right), \quad (j=1,2,\ldots,p), \\
	\xi | \cdot &\sim& \mathcal{IG} \left(1, 1  + \frac{1}{\tau^2}\right). 
\end{eqnarray*}
It is interesting to note that the conditional posterior distributions for all parameters, except the regression coefficients, are inverse-Gamma, for which efficient samplers exist.
\subsection{Horseshoe logistic regression}
The Gibbs sampling approach proposed in this paper can be extended to other models and other prior distributions. In logistic regression, equation (\ref{eqn:y}) in the hierarchy becomes 
\begin{equation}
	y_i | {\bf x}_i, \bm{\beta} \sim {\rm Binom}(1, 1/(1 + e^{-\psi_i})), \nonumber
\end{equation}
where $y_i \in \{0,1\}$ and $\psi_i = {\bf x}^{\T} \bm{\beta}$ are the log-odds of success. 

Bayesian logistic regression with the horseshoe hierarchy may be implemented using a scale mixture based on $z$-distributions~\cite{GramacyPolson12} or the P\'{o}lya-gamma data augmentation framework~\cite{PolsonEtAl2013} for modelling the logistic function at the top level of the hierarchy. Using the P\'{o}lya-gamma approach, the conditional posterior distribution of $\bm{\beta}$ given auxiliary variables $\bm{\omega} = (\omega_1, \ldots, \omega_n)^{\T}$ and data $y_i$ $(i=1,\ldots,n)$ is a multivariate Gaussian given by
\begin{equation}
	\bm{\beta} | \bm{\omega}, {\bf y} \propto \pi_{\beta}(\bm{\beta}) \exp\left( -\frac{1}{2} ({\bf z} - {\bf X} \bm{\beta})^{\T} \Omega ({\bf z} - {\bf X} \bm{\beta}) \right), \nonumber
\end{equation}
where ${\bf z} = (\kappa_1/\omega_1, \ldots, \kappa_n/\omega_n)^{\T}$, $\kappa_i = y_i - 1/2$, $\bm{\Omega} = {\rm diag}(\bm{\omega})$ and $\pi_{\beta}(\bm{\beta})$ is a prior distribution for the regression coefficients. The conditional posterior distribution for the auxiliary variables $\bm{\omega}$ is
\begin{equation}
	\omega_i | \bm{\beta} \sim {\rm PG}(1, {\bf x}_i^{\T} \bm{\beta}), \nonumber
\end{equation}
which is a P\'{o}lya-gamma distribution with shape parameter $1$ and scale parameter ${\bf x}_i^{\T} \bm{\beta}$. With the horseshoe prior, the conditional posterior distribution for the regression coefficients becomes
\begin{equation}
\label{eqn:logreg:beta}
	\bm{\beta} | \cdot \sim \mathcal{N}_p ( {\bf A}^{-1} {\bf X}^{\T} \bm{\Omega} {\bf z}, {\bf A}^{-1} ), \quad {\bf A} = ({\bf X}^{\T} \bm{\Omega} {\bf X} + \bm{\Lambda}_*^{-1}), \quad \bm{\Lambda}_* = \tau^2 \bm{\Lambda} 
\end{equation}
where $\bm{\Lambda} = {\rm diag}(\lambda_1^2,\ldots,\lambda_p^2)$. The conditional posterior distributions for the prior hyperparameters remain unchanged. Unlike in the case of linear regression, the intercept parameter must now be explicitly modelled. Special care should be taken to ensure that the intercept parameter is not penalized (e.g., by using a uniform prior distribution). 

\subsection{Horseshoe negative-binomial regression}
The P\'{o}lya-gamma data augmentation strategy may also be used to derive a Bayesian horseshoe estimator of negative-binomial regression for count data~\cite{PolsonEtAl2013}. Data $y_i$ ($i=1,2,\ldots,n$) is now assumed to be generated by the negative binomial distribution
\begin{equation}
	p(y_i | h, \pi_i) \propto (1 - \pi_i)^h \pi_i^{y_i}, (h > 0)
\end{equation}
where $\pi_i =\exp(\psi_i) / (1 + \exp(\psi_i))$ and $\psi_i = {\bf x}^{\T} \bm{\beta}$. This is equivalent to the sampling model
\begin{equation}
	{\bf z} | {\bf X}, \bm{\beta} \sim \mathcal{N}_n ({\bf X} \bm{\beta}, \bm{\Omega^{-1}})
\end{equation}
where $z_i = (y_i - h) / 2$ and $\bm{\Omega} = {\rm diag}(\omega_1, \omega_2, \ldots, \omega_n)$. The conditional posterior distribution for the auxiliary variables $\bm{\omega}$ is now
\begin{equation}
	\omega_i | \bm{\beta} \sim {\rm PG}(y_i + h, {\bf x}_i^{\T} \bm{\beta}), \nonumber
\end{equation}
which is a P\'{o}lya-gamma distribution with shape parameter $(y_i + h)$ and scale parameter ${\bf x}_i^{\T} \bm{\beta}$. The conditional posterior distribution of the regression coefficients $\bm{\beta}$ with the horseshoe prior is equivalent to the horseshoe logistic regression model (\ref{eqn:logreg:beta}). As before, the intercept parameter must be explicitly modelled and should not be subject to any shrinkage.

\subsection{Alternative Prior Distributions}
Extensions of the hierarchy (\ref{eqn:y})--(\ref{eqn:sigma2}) to other prior distributions, such as the horseshoe$+$~\cite{BhadraEtAl15}, are straightforward and involve a direct application of the auxiliary variable representation. In the horseshoe$+$ hierarchy, the local shrinkage parameters $\lambda_j$ are given the half-Cauchy prior distribution
\begin{equation}
		\lambda_j  \sim \mathcal{C}^+(0, \eta_j^2)
\end{equation}
where $j=(1,2,\ldots,p)$ and $\eta_j$ is a further half-Cauchy mixing variable which is given the prior density
\begin{equation}
		\eta_j^2  \sim \mathcal{C}^+(0, 1).
\end{equation}
The conditional posterior distribution of the local-shrinkage coefficients $\lambda_j$ is equivalent to (\ref{eqn:hs:lam2}) in the case of the horseshoe$+$ estimator. The conditional posterior density for the auxiliary variables $\nu_j$ is now
\begin{equation}
	\nu_j | \cdot \sim \mathcal{IG} \left(1, \eta_j^2 + \frac{1}{\lambda_j^2} \right), \quad (j=1,2,\ldots,p), \\
\end{equation}
Sampling of the auxiliary mixing variables $\eta_j$ is then a straightforward application of the decomposition (\ref{eqn:half-cauchy}).

\begin{table*}[t]
\scriptsize
\begin{center}
\begin{tabular}{lrccccc} \toprule[1pt]
  & Sample size ($n$)   & \multicolumn{5}{c}{Number of predictors ($p$)} \\ 
\cmidrule{3-7}
            &         & ~~~10~~~      & ~~~50~~~     & ~~~100~~~    & ~~~500~~~   & ~~~1000~~~ \\
\cmidrule{1-7}
					&    10 &  0.44 &  0.34 &  0.43 &  1.94 &  5.96 \\
					&    50 &  0.21 &  0.28 &  0.45 &  2.31 &  6.76 \\
{\tt bhs}	&   100 &  0.22 &  0.29 &  0.57 &  2.99 &  7.50 \\
					&   500 &  0.24 &  0.44 &  0.65 &  8.36 & 32.46 \\
					&  1000 &  0.30 &  0.53 &  0.83 & 10.05 & 34.75 \\
\\
							&    10 &  0.01 &  0.23 &  1.49 &165.21 &1470.74 \\
							&    50 &  0.02 &  0.25 &  1.51 &165.06 &1480.70 \\
{\tt monomvn}	&   100 &  0.02 &  0.24 &  1.48 &165.29 &1474.30 \\
							&   500 &  0.05 &  0.31 &  1.57 &165.49 &1488.92 \\
							&  1000 &  0.09 &  0.41 &  1.73 &166.27 &1481.17 \\
  \bottomrule[1pt]
\end{tabular}
\vspace{2mm}
\caption{\small Comparison of run times between the R package {\tt monomvn} and our implementation ({\tt bhs}) of the Bayesian horseshoe sampler.  In each test, the samplers generated $1,000$ samples from the posterior distribution of the Bayesian horseshoe. All timings are given in seconds. \label{tab:TimingResults}}
\end{center}
\end{table*}

\section{Evaluation}
\begin{figure}
\begin{center}
\includegraphics[width=12cm]{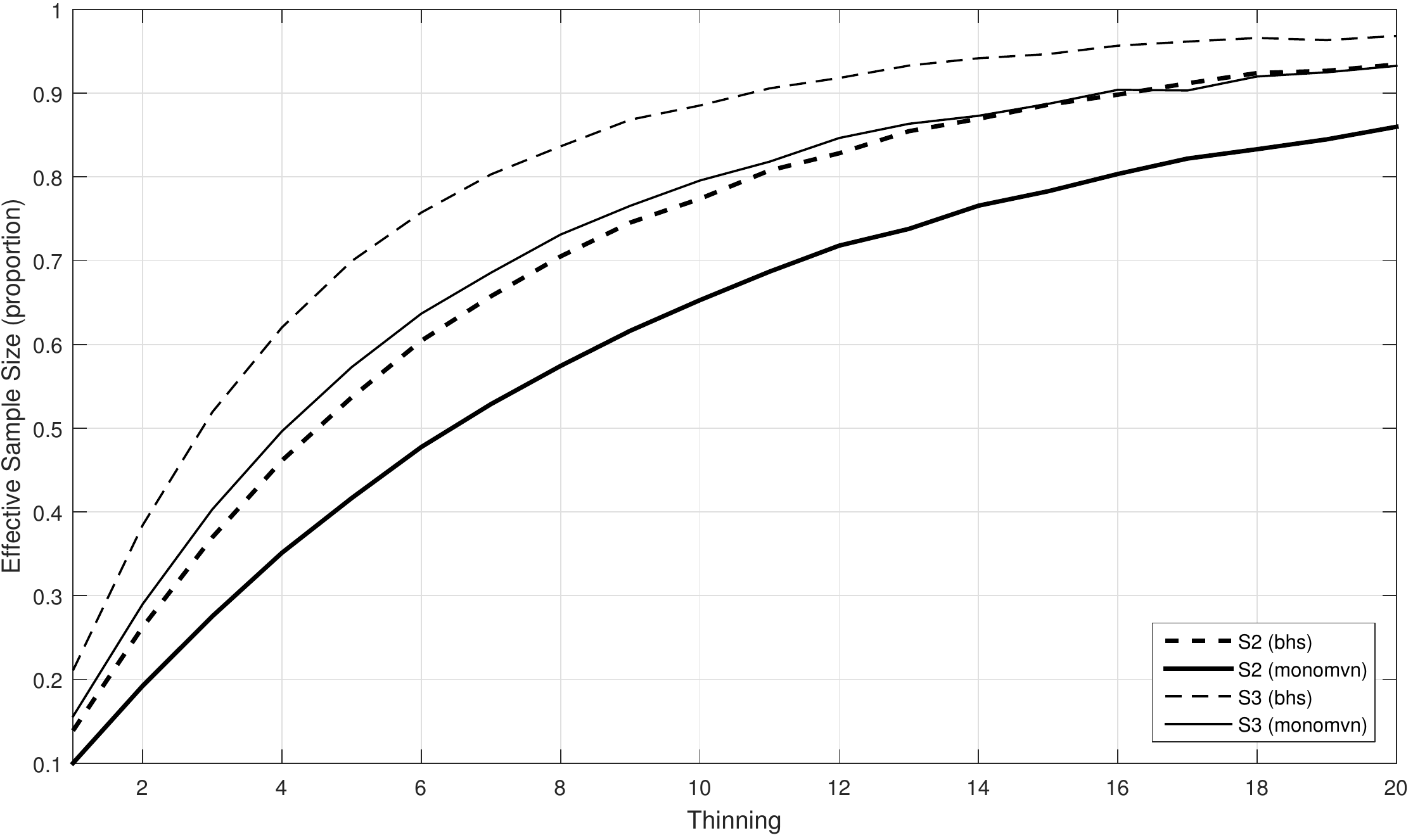}
\caption{Comparison of sampling efficiency between {\tt monomvn} and {\tt bhs} in terms of effective sample size (expressed as a proportion)}
\label{fig:ess:comparison}
\end{center}
\end{figure}
We have conducted a simulation experiment to compare run times between our implementation ({\tt bhs}) and the R package {\tt monomvn} implementation (available on the CRAN repository) of the Bayesian horseshoe estimator. This comparison is expected to be somewhat biased as the two packages are implemented in different programming languages --- our implementation is in pure MATLAB code, while the {\tt monomvn} package is implemented as an R interface to compiled C code. The execution times of both procedures were compared across a range of values for sample size and the number of predictors. For each timing run, both procedures were timed generating $1,000$ samples from the Bayesian horseshoe posterior distribution and all timing results were measured in seconds. The results of the timing experiment are given in Table~\ref{tab:TimingResults}. For small sample sizes and small numbers of predictors, the {\tt monomvn} implementation is faster than {\tt bhs}, with both procedures finishing in less than one second of execution time. However, as the number of predictors grows our implementation is significantly faster than {\tt monomvn}. As an example, for the experiment ($n=1,000$, $p=1,000$), our implementation is approximately $40$ times faster than {\tt monomvn}. 

The {\tt monomvn} package uses the conventional approach to sampling from the posterior distribution of the horseshoe estimator by slice sampling the hyperparameters $\tau$ and $(\lambda_1, \ldots, \lambda_p)$. We compared our sampling procedure based on auxiliary variables ({\tt bhs}) against the {\tt monomvn} implementation using the diabetes data set examined in~\cite{ParkCasella08}. This dataset consists of 442 observations and 10 predictors, some of which are highly correlated. 

The sampling efficiency of the two procedures was compared using the effective sample size metric as discussed in~\cite{Geyer92}. The effective sample size for each regression coefficient was estimated, under varying levels of thinning, from a chain of $50$ million posterior samples generated by both {\tt bhs} and {\tt monomvn}. For all regression coefficients and under all levels of thinning, the {\tt bhs} procedure was found to have a higher effective sample size. As an example, Figure~\ref{fig:ess:comparison} depicts the effective sample size for predictors {\tt S2} and {\tt S3} as a function of the level of thinning. These two predictors had the smallest observed effective sample size of the ten predictors in the diabetes data set. The {\tt bhs} procedure is clearly more efficient than {\tt monomvn} and the difference appears greatest for thinning levels of four to eight. In order to achieve an effective sample size of $80\%$ for all ten predictors, the {\tt monomvn} sampler required a thinning level of 16; in contrast, the {\tt bhs} sampler required a thinning level of 11 to achieve the same sampling efficiency. Despite the introduction of latent variables, the {\tt bhs} algorithm appears to be more efficient in terms of effective sample size than the {\tt monomvn} procedure which is based on slice sampling with no auxiliary variables. This may be due to the slice sampler suffering from numerical issues driven by the heavy tails of the Cauchy prior distributions. 


\section{Discussion}
%
%
%

The decomposition (\ref{eqn:half-cauchy}) reveals an interesting novel interpretation of the horseshoe hierarchy. Integrating out the hypervariances $(\lambda_1^2, \ldots, \lambda^2_p)$ implies a Cauchy prior distribution over each regression coefficient $\beta_j$ of the form
%
\begin{equation}
		\beta_j | \tau^2, \sigma^2, \nu_j \sim \mathcal{C}(0, 2 \tau \sigma / (2 \nu_j)^{1/2}). \nonumber
\end{equation}
The horseshoe model can therefore be viewed as placing a Cauchy prior distribution over each regression coefficient $\beta_j$ with the scale of each prior inversely proportional to $\nu_j$. 

Gibbs sampling may also be used in the case where the global scale parameter $\sigma > 0$ is given a half-Cauchy prior distribution, as recommended in~\cite{PolsonScott12}, by using the decomposition (\ref{eqn:half-cauchy}). An interesting consequence of the conjugacy of all conditional posterior distributions is that Chib's algorithm~\cite{Chib95} for computing the marginal likelihood from the output of a Gibbs sampler is readily applicable as the normalizing constants for all conditionals are known. This is in contrast to sampling methods that use algorithms such as slice sampling.

Our MATLAB\texttrademark{} implementation of Bayesian horseshoe linear regression is available from the MATLAB Central File Exchange repository (File ID \#52479). The implementation uses Rue's algorithm~\cite{Rue01} for efficient sampling from the multivariate Gaussian conditional posterior distribution of the regression coefficients when the sample size is greater than the number of predictors ($n > p$). This algorithm is based on Cholesky factorization with cubic order of complexity in terms of the number of predictors $p$. For the case where the number of predictors is large, this approach for sampling from multivariate Gaussian distributions is computationally inefficient. In this setting, our implementation employs the sampling algorithm given in~\cite{BhattacharyaEtAl15} which has linear complexity in terms of the number of predictors.


\bibliographystyle{unsrt}
\bibliography{bibliography}

\begin{thebibliography}{10}

\bibitem{CarvalhoPolson10}
Carlos~M. Carvalho, Nicholas~G. Polson, and James~G. Scott.
\newblock The horseshoe estimator for sparse signals.
\newblock {\em Biometrika}, 97(2):465--480, 2010.

\bibitem{ParkCasella08}
Trevor Park and George Casella.
\newblock The {B}ayesian lasso.
\newblock {\em Journal of the American Statistical Association},
  103(482):681--686, June 2008.

\bibitem{PolsonScott12a}
Nicholas~G. Polson and James~G. Scott.
\newblock Local shrinkage rules, {L}\'{e}vy processes and regularized
  regression.
\newblock {\em Journal of the Royal Statistical Society (Series B)},
  74(2):287--311, 2012.

\bibitem{PolsonScott10}
Nicholas~G. Polson and James~G. Scott.
\newblock Shrink globally, act locally: Sparse {B}ayesian regularization and
  prediction.
\newblock In {\em Bayesian Statistics}, volume~9, 2010.

\bibitem{Neal03}
Radford~M. Neal.
\newblock Slice sampling.
\newblock {\em The Annals of Statistics}, 31(3):705--741, June 2003.

\bibitem{PolsonEtAl2014}
Nicholas~G. Polson, James~G. Scott, and Jesse Windle.
\newblock The {B}ayesian bridge.
\newblock {\em Journal of the Royal Statistical Society (Series B)},
  76(4):713--733, 2014.

\bibitem{WandEtAl11}
Matthew~P. Wand, John~T. Ormerod, Simone~A. Padoan, and Rudolf Fruhwirth.
\newblock Mean field variational {B}ayes for elaborate distributions.
\newblock {\em Bayesian Analysis}, 6(4):847--900, 2011.

\bibitem{LindleySmith72}
D.~V. Lindley and A.~F.~M. Smith.
\newblock Bayes estimates for the linear model.
\newblock {\em Journal of the Royal Statistical Society (Series B)},
  34(1):1--41, 1972.

\bibitem{GramacyPolson12}
Robert~B. Gramacy and Nicholas~G. Polson.
\newblock Simulation-based regularized logistic regression.
\newblock {\em Bayesian Analysis}, 7(3):567--590, 2012.

\bibitem{PolsonEtAl2013}
Nicholas~G. Polson, James~G. Scott, and Jesse Windle.
\newblock Bayesian inference for logistic models using {P}\'{o}lya-gamma latent
  variables.
\newblock {\em Journal of the American Statistical Association},
  108(504):1339--1349, 2013.

\bibitem{BhadraEtAl15}
Anindya Bhadra, Jyotishka Datta, Nicholas~G. Polson, and Brandon Willard.
\newblock The horseshoe+ estimator of ultra-sparse signals.
\newblock 2015.
\newblock arXiv:1502.00560.

\bibitem{Geyer92}
Charles~J. Geyer.
\newblock Practical markov chain monte carlo.
\newblock {\em Statistical Science}, 7(4):473--483, 1992.

\bibitem{PolsonScott12}
Nicholas~G. Polson and James~G. Scott.
\newblock On the half-{C}auchy prior for a global scale parameter.
\newblock {\em Bayesian Analysis}, 7(4):887--902, 2012.

\bibitem{Chib95}
Siddhartha Chib.
\newblock Marginal likelihood from the {G}ibbs output.
\newblock {\em Journal of the American Statistical Association},
  90(432):1313--1321, December 1995.

\bibitem{Rue01}
H.~Rue.
\newblock Fast sampling of gaussian markov random fields.
\newblock {\em Journal of the Royal Statistical Society (Series B)},
  63(2):325--338, 2001.

\bibitem{BhattacharyaEtAl15}
Anirban Bhattacharya, Antik Chakraborty, and Bani~K. Mallick.
\newblock Fast sampling with {G}aussian scale-mixture priors in
  high-dimensional regression.
\newblock 2015.
\newblock arXiv:1506.04778.

\end{thebibliography}

\end{document}